\documentclass[notitlepage]{article}
\usepackage{graphicx} 
\usepackage{amssymb,amsfonts,amsmath}
\usepackage{txfonts}
\usepackage[font=small]{caption} 
\usepackage{xcolor}              
\usepackage{setspace}            
\usepackage{authblk}             
\usepackage{fullpage}            
\usepackage{url}
\usepackage{caption}
\usepackage{subcaption}
\usepackage{booktabs}
\usepackage{url}
\usepackage{cite}	
\def\etal{{et~al.}}
\def\lett#1{(\textbf{#1})}

\definecolor{darkgreen}{HTML}{00BB00}

\def\perc{\mathrm{perc}}
\def\GCC{\mathrm{GCC}}
\def\CV{\operatorname{CV}}
\newcommand{\avg}[1]{\ensuremath{\left<#1\right>}}

\author[1,$^*$]{Louis M.~Shekhtman}
\author[2,3]{James P.~Bagrow}
\author[2]{Dirk ~Brockmann}
\affil[1]{\small Department of Physics and Astronomy, Northwestern University, Evanston, IL 60208, USA}
\affil[2]{Department of Engineering Sciences and Applied Mathematics, Northwestern Institute on Complex Systems, Northwestern University, Evanston, IL 60208, USA}
\affil[3]{Department of Mathematics and Statistics, Vermont Advanced Computing Center, Complex Systems Center, University of Vermont, Burlington, VT 05405, USA}
\affil[$^*$]{Corresponding author: lshek@u.northwestern.edu}

\title{Robustness of skeletons and salient features in networks}

\begin{document}
\maketitle

\begin{abstract} 
    Real world network datasets often contain a wealth of complex topological
    information. In the face of these data, researchers often employ methods to
    extract reduced networks containing the most important structures or
    pathways, sometimes known as `skeletons' or `backbones'. Numerous such
    methods have been developed. 
    Yet data are often noisy or incomplete, with unknown numbers of missing or
    spurious links. Relatively little effort has gone into understanding how
    salient network extraction methods perform in the face of noisy or
    incomplete networks.
    We study this problem by comparing how the salient features extracted by two
    popular methods change when networks are perturbed, either by deleting nodes
    or links, or by randomly rewiring links. 
    Our results indicate that simple, global statistics for skeletons can be
    accurately inferred even for noisy and incomplete network data, but it is
    crucial to have complete, reliable data to use the exact topologies of
    skeletons or backbones. 
    These results also help us understand how skeletons respond to damage to the
    network itself, as in an attack scenario. 
\end{abstract}

\section{Introduction}
Many systems consist of discrete elements that are coupled to one another in sophisticated ways. 
Modeling these systems as networks often exposes more clearly the fundamental properties of
 the dataset \cite{network-basics2, network-motifs, newman1, network-basics}. While modeling 
systems as networks is not a new approach, it has become more prevalent due to the greater
 availability of large data sets \cite{motter-networks}. The brain's neurons have been mapped
 using these methods \cite{neural-networks}, as have air traffic patterns \cite{air-transportation-networks},
 and the flow of cargo throughout the world \cite{global-cargo-network}.

The explosion of research on complex networks in recent years has led to the
discovery of various properties of networks and has allowed us to find ways of
reducing the complexity while preserving certain key features. 
Many of these methods focus on reducing the number of nodes in the network. 
Aside from simple thresholding, more sophisticated coarse-graining techniques have also been used \cite{coarse-graining} to reduce the number of distinct entities in the network. Here we will focus on methods of reducing the number of links in the network while preserving the nodes. This is advantageous since it reduces the complexity of the system while still preserving scale-free properties.

Further, there
has been considerable effort in understanding how networks as a whole respond to
damage\cite{barabasi-attack-tolerance, havlin-attack-tolerance, newman-percolation,watts-percolation}. These studies have explored different methods of perturbing the network such as intentional attack and random failure. 

Despite the significant amount of research in both of these areas separately, there has been little work in combining the study of backbone and skeleton methods with stress applied to the system. Here we examine how  skeletons and backbones respond to different methods of stress applied to the system.

\subsection{Network data}
In exploring the response of network skeletons to perturbations to the network
as a whole, we use three different transportation networks, three biological
networks and one network model. The transportation networks used are the world
air transportation network from $1995$ (Airport), the network of global cargo
shipments (Cargo), and the network of human migrations provided by the IRS
(Migration). The Airport network was taken from OAG Worldwide Ltd. and has been examined in various previous studies \cite{airport1, air-transportation-networks, airport2}. The Cargo network comes from the IHS Fairplay data and contains information about $16,323$ container ships \cite{global-cargo-network}. 

 For biological networks we examine the network containing the
neural interactions of C. elegans (Neural), the Florida Bay food web (Food Web),
and the metabolic network of \emph{E. coli} (Metabolic). The Neural network comes from work by White \etal \cite{neural-data} and was explored in \cite{neural-analysis}.  The Food Web is from a collection of public data sets available online \cite{foodweb1}. Finally, the Metabolic network comes from experimental research and has also been previously analyzed \cite{metabolic1}.

 Lastly, we analyze an
Erd\"{o}s-R\'{e}nyi network with link weights drawn from a power-law distribution
(Random).  Basic summary statistics for the networks, such as the number of
nodes $N$ and links $L$, is provided in Table~\ref{table:BasicInfo}.

\begin{table}
    \small
\centering
\begin{tabular}{rccccccc}
\toprule
Network       & $N$    & $L$     & $\avg{k}$   & $\rho$  & $\CV(k)$ & $\CV(w)$ & $r$\\
\midrule
Airport       & $1227$ & $18050$ & $29.42$ & $0.024$ & $1.29$  & $2.25$  & $-0.06$ \\ 
Migration     & $3056$ & $71551$ & $46.83$ & $0.015$ & $1.94$  & $6.13$  & $-0.06$ \\ 
Cargo         & $951$  & $25819$ & $54.30$ & $0.057$ & $1.22$  & $ 6.85$ & $-0.14$ \\ 
Neural        & $297$  & $2141$  & $14.46$ & $0.049$ & $0.89$  & $1.35$  & $-0.16$ \\ 
Food Web      & $121$  & $1763$  & $29.14$ & $0.24$  & $0.45$  & $11.77$ & $-0.10$ \\ 
Metabolic     & $311$  & $1304$  & $8.39$  & $0.027$ & $1.79$  & $7.91$  & $-0.25$ \\ 
Random        & $1000$ & $15028$ & $30.06$ & $0.03$  & $0.177$ & $1.44$  & $-0.005$ \\
\bottomrule
\end{tabular}
\caption{Summary of the networks. Presented here: $N$, the number of nodes in
    the network; $L$, the number of links; $\avg{k}$, the average degree; $\CV(k)$ the
cofficient of variation of degree; $\CV(w)$, the coefficient of variation of
weight; $\rho = L / \binom{N}{2}$, the network density; and $r$, the degree assortativity coefficient.}
\label{table:BasicInfo}
\end{table}

\subsection{Skeleton methods}

While there are many ways to extract the most central links,
the two methods explored here are the {\bf salience skeleton} of Grady
\etal{}~\cite{salience} and the {\bf disparity backbone} of Serrano
\etal{}~\cite{backbone}. Both of these methods involve using the weights on the
network links and therefore require that the data be presented as a weighted
network.  Note that while the terms 'backbone' and 'skeleton' are generally 
synonymous, for clarity we will refer to the 
\emph{salience skeleton} and \emph{disparity backbone} for these methods. 

\begin{figure}
\centering
        \includegraphics[width=.5\linewidth]{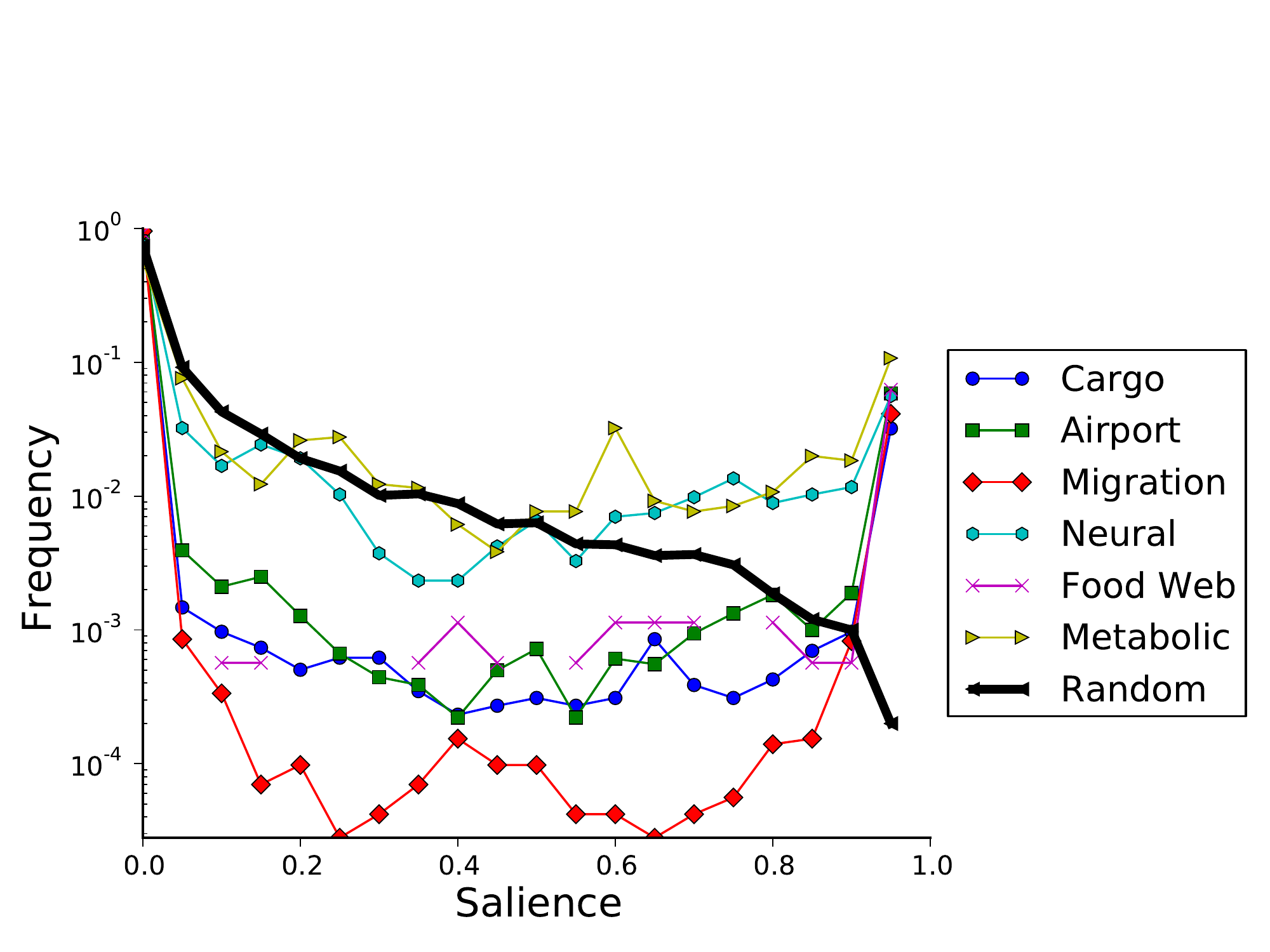}
\caption{\onehalfspacing The salience distribution of the networks
used in the analysis.  For all the networks which use real data the distribution
displays a bimodal characteristic. Notably, the curve of the random network does
not have the same shape. Values not shown are at zero.}
    \label{fig:salience-dist}
\end{figure}

The salience skeleton is an analysis based on the shortest path trees (SPTs) of a
network and is similar to the method used by Wu \etal{}~to find superhighways
\cite{havlin-transport}. First we compute the SPT rooted at each node of the
network using Dijkstra's algorithm. The \textbf{salience} $S_{ij}$ of edge $ij$ is then
the fraction of
SPTs in which $ij$ appears~\cite{salience}:
\begin{equation}
    S_{ij}=\frac{1}{N} \sum_{c=1}^{N} \Big[(ij) \in T_c \Big]
    \label{eqn:salience}
\end{equation}
where $T_c$ is the set of all edges in the SPT rooted at node $c$ and
$\left[P\right] = 1$ if statement $P$ is true and zero otherwise.
In real networks salience is distributed
bimodally (Fig.~\ref{fig:salience-dist}), meaning that links occur in nearly all
SPTs ($S\approx 1)$ or in almost none ($S\approx 0)$. This makes it a natural way
of extracting a network skeleton without having to choose an arbitrary cutoff
for $S$.
Note that Eq.~\eqref{eqn:salience} is very similar to edge betweenness but subtly
distinct in that it counts each \emph{tree} whereas betweenness counts each 
\emph{path}~\cite{salience}.

The disparity backbone method focuses on statistically significant deviations in
link weight. One begins by defining a null model that determines the
expected distribution of link weights around a node with $k$ links, if those weights
were distributed randomly between the links. The method then compares the actual
link weights around the node to the null model. A significance level $\alpha \in
(0,1)$ is chosen
and all links that are statistically significant at $\alpha$ belong to the
disparity backbone~\cite{backbone}.

\subsection{Robustness methods}

In perturbing the networks we explore (i) node percolation, (ii) link
percolation and (iii) link switching. 
We define the percolation either of links or nodes by the number $p_{\perc}$
which is the fraction of links or nodes removed from the network. The classic
result from percolation involves a phase transition in the size of the giant
connected component (GCC) for random networks. For most real networks
there is no phase transition (while $p_{\perc} < 1$) and the size of the giant connected component is
robust. We repeat this experiment and examine how the giant connected component
changes under link percolation for our datasets. We confirm the previous
results, which have shown that real networks are robust to link percolation. 

\begin{figure}
\centering
        \includegraphics[width=.5\linewidth]{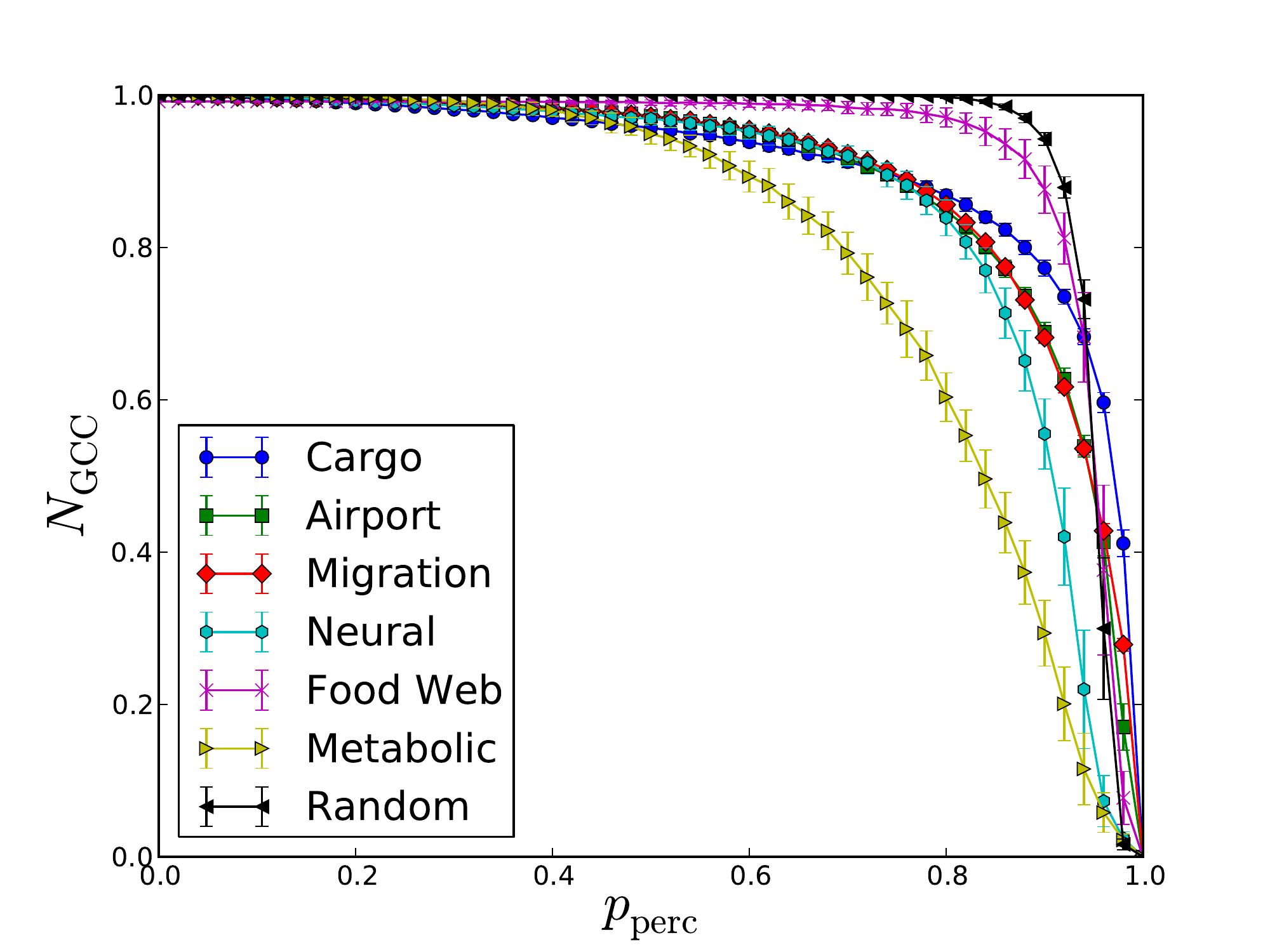}
\caption{\onehalfspacing The giant connected component under link percolation.
    We observe that the fraction of nodes in the giant connected component,
    $N_{\GCC}$, of the network is robust under link percolation meaning most
    links must be removed before the network fragments. This holds for all
networks studied here.}
    \label{fig:link-perc_ngcc}
\end{figure}

There are a variety of methods of
performing link rewiring and the process is somewhat subtle. We use the method
introduced by Karrer \etal{}~\cite{PhysRevE.77.046119},   which involves rewiring
in such a way that the expectation value of the degree of
each node is preserved. This is done by defining the probability of an edge,
$e_{ij}$, existing
between nodes $i$ and $j$ according to their degrees:
\begin{equation}
    e_{ij}=\frac{k_i k_j}{2L},
\end{equation}
where $k_i$ is the degree of node $i$.
To rewire, we go through each edge in the network and with some
probability $p_s$ we remove that edge and insert a new edge between nodes $i$
and $j$, with $i$ and $j$ chosen with probability $e_{ij}/L$. Otherwise, with
probability $1-p_s$, we leave that original edge in place.
Karrer \etal{}~show that this rewiring scheme preserves the expected degree of each node in the
network while allowing us to tune the quantity of randomness with the parameter
$p_s$.

\section{Results}

We now study how our skeleton methods perform in the face of noisy and missing
data by applying them to perturbed versions of our networks and comparing their
results to those obtained for the original networks.

\begin{figure}
    \centering
\includegraphics[width=.5\textwidth]{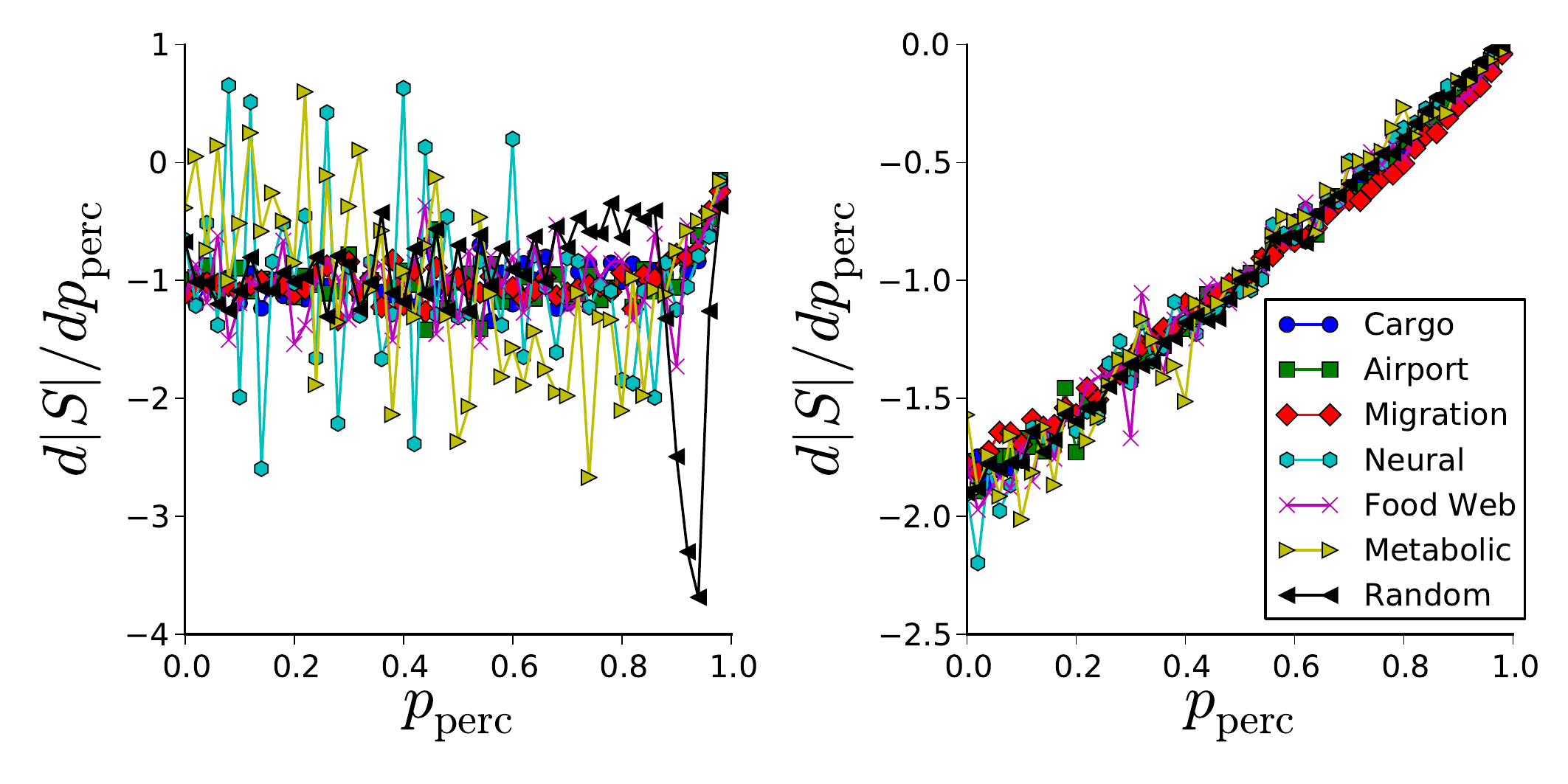}
\caption{\onehalfspacing
       How site (or node) percolation changes the size of the \lett{left} salience skeleton
       and the \lett{right} disparity backbone.
       The linear decrease in the size of the skeleton skeleton as $p_{\perc}$
       increases shows that the size of the skeleton is proportional to $N$, the
       number of nodes in the network.  (The noise in the Neural and Metabolic
       networks is likely due to the smaller size of those networks.)
       Meanwhile, the disparity backbone decreases in size more quickly than
       the salience skeleton for $p_\perc < 1/2$. 
       The disparity
       backbone is more sensitive to site percolation than the salience skeleton, especially 
       for small amounts of percolation.
       \label{fig:figure-site-perc}
}
\end{figure}

In the case of node percolation we observe that the size or fraction of links in
the skeleton, $|S|$, is roughly proportional to $N$, the number of nodes in the
network. This can be seen by the fact that $\tfrac{d|S|}{dp_{\perc}}\approx -1$.
For the salience skeleton this is true for all values of $p_{\perc}$ while for
the disparity backbone the linear regime terminates earlier. This is shown in
Fig.~\ref{fig:figure-site-perc}. This suggests that for the salience skeleton it
is mainly the path to the removed node that is affected by the percolation while
paths to other nodes may change slightly but contain about the same number of
links as the original path. For the disparity backbone the decrease is faster
than for the salience skeleton which shows that the size of the disparity
backbone is more sensitive to the number of nodes in the network.

In examining changes to the links in the network we look at several other
quantities. First the skeleton giant connected component, $S_{\GCC}$ is
intuitively defined as the fraction of the network that is connected when the
network is reduced to its skeleton. Second, we examine how many links are added
to the skeleton, $L_A$, after perturbation, and how many links are deleted from
the skeleton, $L_D$, after perturbation. The comparisons are always made to the
original skeleton without any perturbation.

\begin{figure}   
\centering
\includegraphics[width=0.5\linewidth]{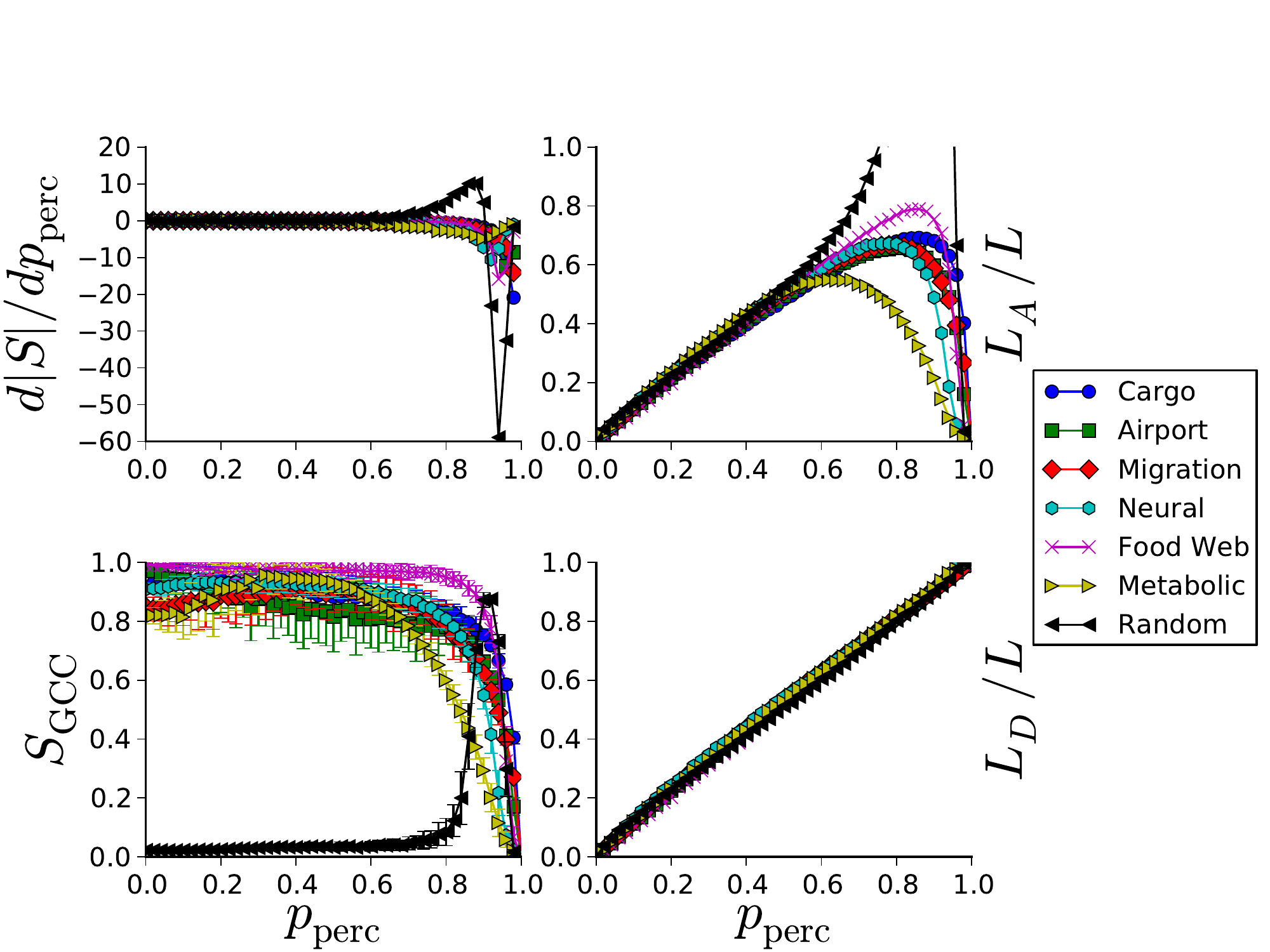}
\caption{\onehalfspacing Changes to the salience skeleton under link percolation.
    \lett{top left} %
    The size of the skeleton is robust to link percolation since
    $\frac{d|S|}{dp_{\perc}}\approx 0$ until $p_{\perc} \rightarrow 1$.
    \lett{bottom left}
    The giant connected component of the skeleton itself is also robust to link
    percolation (except in the case of the random network). This is similar to
    the giant component of the network as a whole which was previously shown to
    be robust to link percolation.
    \lett{top right} %
    Despite the robustness of the size of the skeleton, there are many new links
    that are added to the skeleton as we increase $p_{\perc}$.
    \lett{bottom right} %
    Further we observe that an equivalent number of links are removed from the
    skeleton which leads to the lack of change in its size. This
    demonstrates that the skeleton performs a balancing
    act where  removed links are compensated with new links. New shortest paths
    are found and these new paths contain approximately the same number of links
    as the old ones.
}
\label{fig:salience-link-perc}
\end{figure}

For link percolation we observe in Fig.~\ref{fig:salience-link-perc} that for
the salience skeleton both the size of the skeleton and the size of the skeleton
giant connected component ($S_{\GCC}$) are robust to change. However, the
plots of $L_A$ and $L_D$ make clear that the salience skeleton itself is
undergoing significant changes. Essentially this suggests that under link
percolation the salience skeleton is able to find replacement pathways and those
paths are not considerably longer than the original paths. Links are being
added and deleted, yet the skeleton is simply rerouted and maintains its
connectivity and size.

\begin{figure}
\centering
        \includegraphics[width=0.5\linewidth]{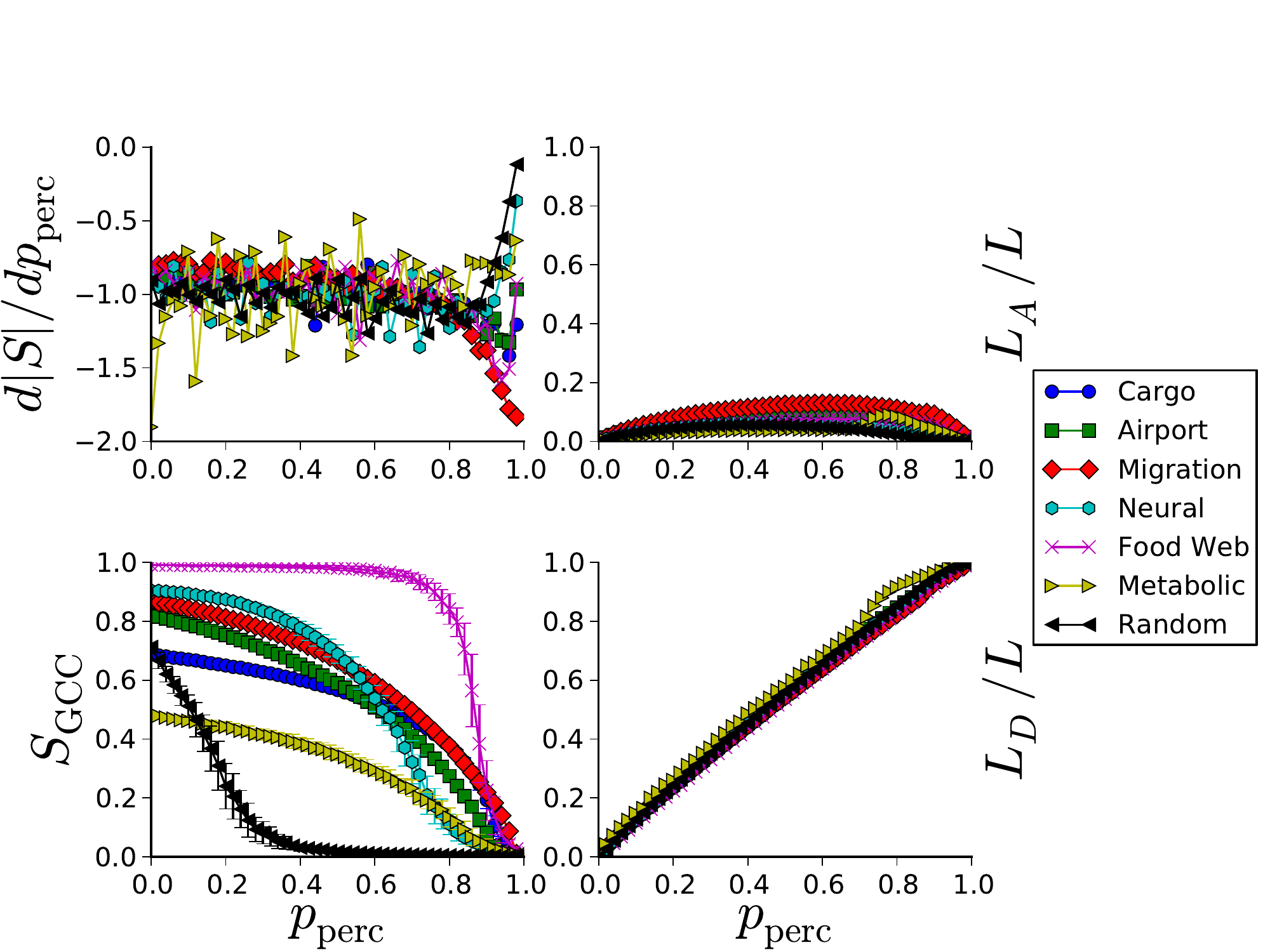}
    \caption{\onehalfspacing Changes to the disparity backbone under
        link percolation.
        \lett{top left} %
        The size of the backbone is not robust to link percolation which is in
        contrast to the salience skeleton. 
	\lett{bottom left} %
In agreement with the salience skeleton, the giant connected component of
the disparity backbone is also robust to link percolation, yet not quite to
the same extent. 
        \lett{top right} %
        Relatively few links are added to the disparity backbone as links
        are deleted. Again, this is in contrast to the salience skeleton where
        enough links were added to compensate for the links deleted from the
        skeleton.
        \lett{bottom right} %
    The rate at which links are deleted from the backbone is similar to that for
    the salience skeleton. The main difference however is that this removal of
    links is not compensated for by the addition of new links.
}
  \label{fig:disp-filter-link-perc}
\end{figure}

The one exception to this is the simulated Random network which has a very
fragmented skeleton. This behavior corresponds to the fact that in the Random
network there is a weaker preference for shortest paths, i.e.~the salience is
not bimodal as shown in Fig.~\ref{fig:salience-dist}. However, after we remove a
large fraction of the links each node only has a couple of links and the
shortest paths all go through the same links. 

\begin{figure}
\centering
        \includegraphics[scale=.5]{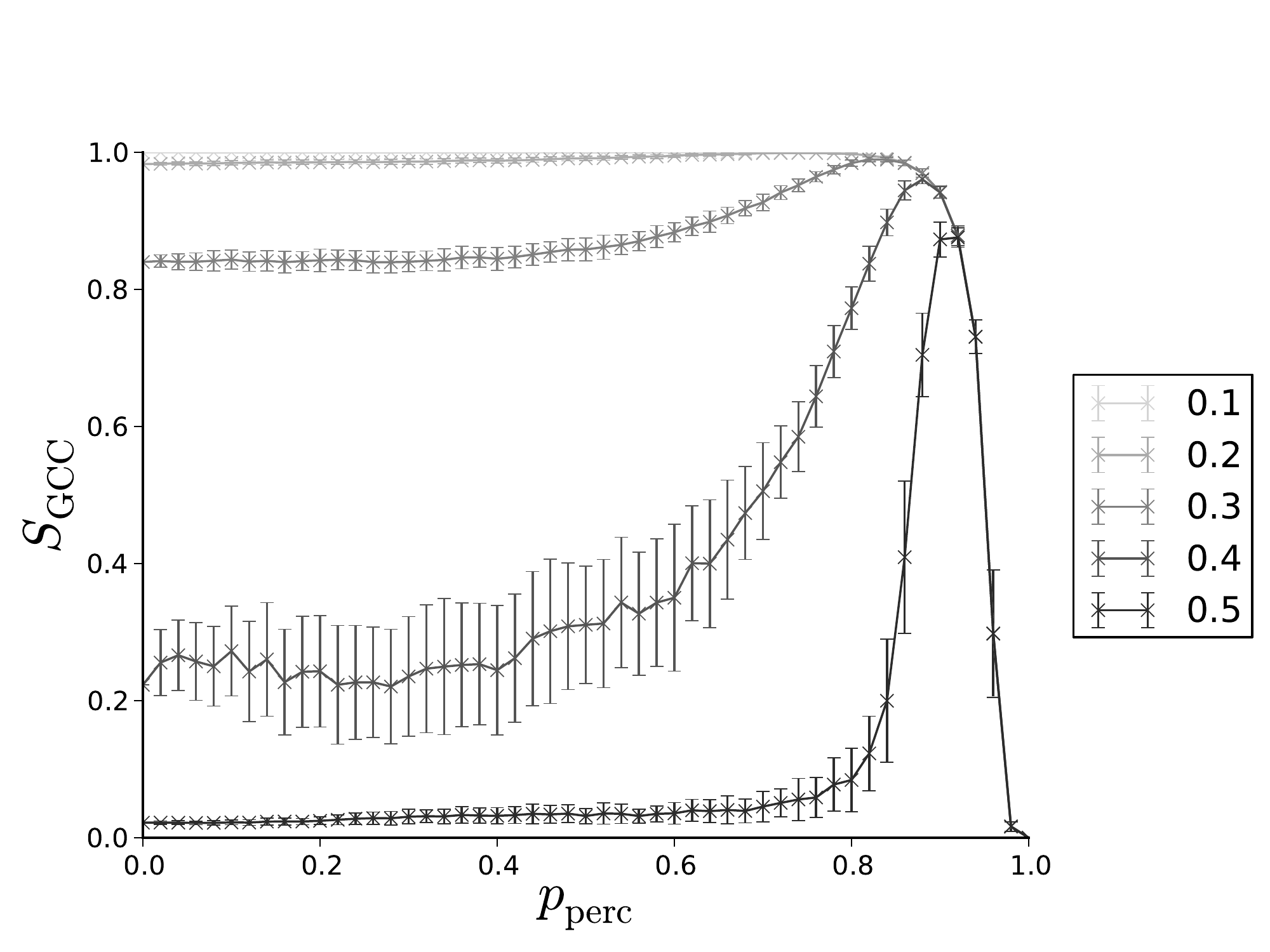}
\captionof{figure}{\onehalfspacing The skeleton giant connected component,
$S_{\GCC}$, of the Random network with different salience cutoffs.  For salience cutoff values $>0.3$ the $S_{\GCC}$ increases at $p_{\perc}\approx .8$ and then begins decreasing at $p_{\perc} \approx .95$.  The unique shape of the salience distribution of the Random network leads to
different cutoffs being required for robustness. } 
\label{fig:er-sal-lim}
\end{figure}

To analyze this hypothesis and confirm that this is not an artifact of the
specific salience cutoff value chosen (0.5), we examine the $S_{\GCC}$ of the
Random network with different salience cutoff values. In
Fig.~\ref{fig:er-sal-lim} we observe that the $S_{\GCC}$ vs. $p_{\perc}$ curve has the same shape until the salience cutoff is very low. At that point the $S_{\GCC}$ of the Random network is also robust to percolation and increasing the amount of percolation never leads to a larger $S_{\GCC}$.   The different
behavior of the Random network shows that real networks have intrinsic
properties which lead the $S_{\GCC}$ to be robust under link percolation. 

Meanwhile, in Fig.~\ref{fig:disp-filter-link-perc} we consider how link
percolation affects the disparity backbone. The backbone size decreases, yet its
giant connected component remains robust. Comparing $L_A$ to $L_D$ shows that
many links are deleted and very few are added to compensate for those removed.
We also observe that the backbone of the Random network is less robust than the
backbones of the real networks, as it was for the salience skeleton.

\begin{figure}
\centering
\includegraphics[width=0.5\linewidth]{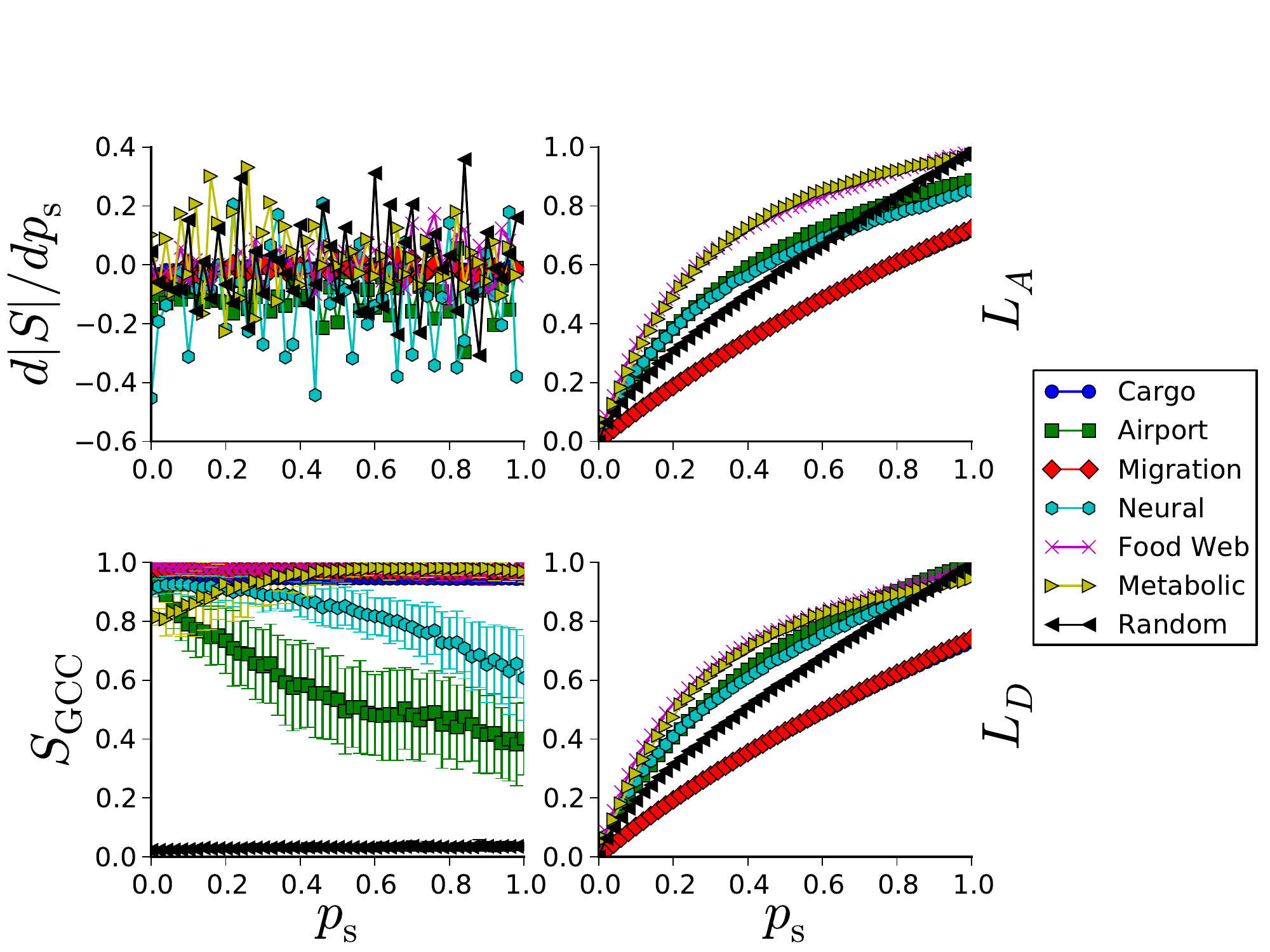}
\caption{\onehalfspacing Changes to the salience skeleton under link switching.
\lett{top left} %
    The size of the skeleton is again robust.
\lett{bottom left} %
    The skeleton giant component of the migration and cargo networks is robust
    to link switching, yet the airport network's skeleton becomes dramatically
    fragmented. This is likely due to a specific, unstable hierarchical (or
    hub-spoke) structure present in the airport network that dictates the paths
    for the salience skeleton. Such a hub-spoke structure may also account for
    the slight decrease in the skeleton size for the neural network.
\lett{top and bottom right}
    Many links are added and removed from the skeleton, once again in a way that
    maintains its size. Further, this reveals that while the size of the
    skeleton can be determined by the number of nodes and the degree
    distribution, knowing which particular links will be present in the skeleton
    requires having the complete dataset.
}
\label{fig:figure-salience-link-switch}
\end{figure}

\begin{figure}
\centering
        \includegraphics[width=0.5\linewidth]{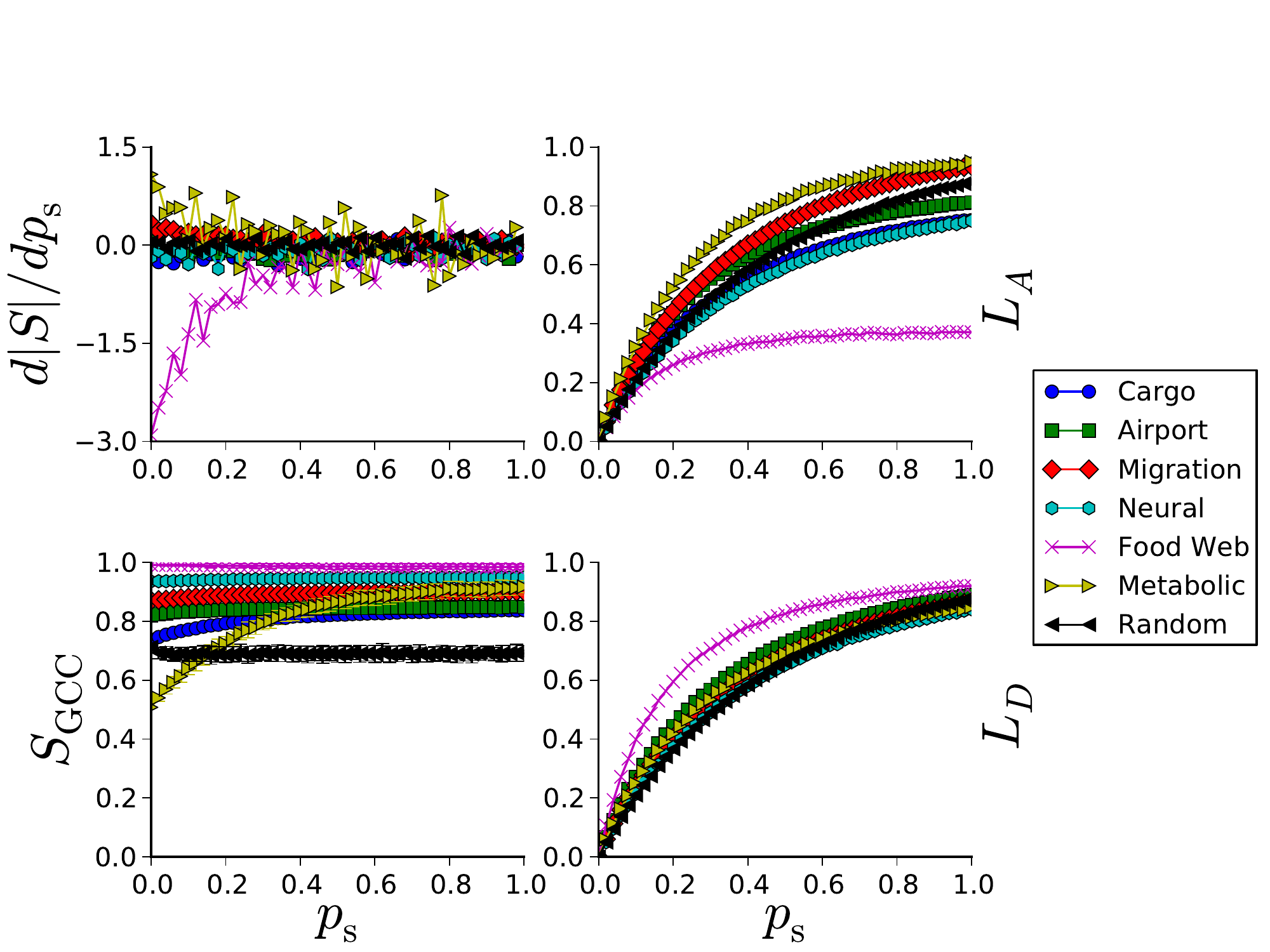}
   
    \captionof{figure}{ \onehalfspacing Changes to the disparity backbone under link switching.
        \lett{top left} %
        As was true for the salience skeleton under link switching and link
        percolation, we find the size of the disparity backbone to be
        robust under link switching.
        \lett{bottom left} %
The giant connected component of the disparity backbone is similarly
robust.
        \lett{top and bottom right} %
    Again there are a significant number of links added and deleted yet they
    once again balance to maintain the size of the skeleton.
}
  \label{fig:figure-disp-filter-link-switch}
\end{figure}

Similarly, upon switching links using the method of
Karrer~\etal{}~\cite{PhysRevE.77.046119}, we observe that $S$ and $S_\GCC$ are
robust for both the salience skeleton
(Fig.~\ref{fig:figure-salience-link-switch}) and the disparity backbone
(Fig.~\ref{fig:figure-disp-filter-link-switch}). 
The significant decrease and large variation in the airport network's salience
skeleton giant component is likely due to a specific, unstable hierarchical
structure present in that network which, when altered, leads to fragmentation.
Further work is needed to determine the exact nature of this structure.  The
neural network exhibits similar behavior likely due to this. The low $S_{\GCC}$
of the random network occurs for the same reason as seen under link percolation.
Once again, we observe that similar to link percolation, despite the robustness
of the skeleton size and giant connected component, there are significant
changes in the links that actually make up the skeleton.  Specifically we see
large changes in $L_A$ and $L_D$ just as we did with link percolation.

\section{Discussion}

These results show that global summary statistics of skeletons, such as the size
of the skeleton and the size of the skeleton giant connected component, are robust to
changes in the network structure. In contrast the specific details of the
skeleton, such as the exact links it contains, will vary, potentially greatly, as the network is
perturbed. This suggests that while skeleton extraction methods are useful for
understanding the global properties of a network, caution should be applied when
attempting to understand local properties based on extraction methods. 

We further showed that different methods of computing the skeleton  respond quite differently under perturbation in
many cases. Lastly, the response of skeletons of real networks is significantly
different than the response of a random network. The methods used to compute
network skeletons and backbones exploit properties of real networks and these
properties are not present in the simulated network. This leads the skeleton of
the random network to respond quite differently under perturbation.

An obvious application of this work is to damage or change in transportation
networks, where skeletons will be responsible for carrying the majority of the system's traffic. This change often occurs in the real world scenario of transport
reroutings and cancellations. The results here show that as these changes occur
the specific composition of the backbone or skeleton changes significantly.
Nonetheless global properties can still often be extracted from the skeleton.

A second application is in protein-protein networks. These networks often
contain noisy data and are  considered incomplete in the interactions  they show
\cite{bayesian-network, yeast-proteins}. Significant work to map
these networks entirely and obtain a full set of all the connections present is ongoing
\cite{yeast-proteins2}. Despite the lack of the full dataset, much analysis has
already been done on the data that is available \cite{yeast-network-analysis,
yeast-network-analysis2}. Our results suggests that caution should be applied
when looking at structural skeletons or backbones for many biological data sets
that contain noisy data because the errors will have a profound impact on the
resulting skeleton and backbone structures.

Lastly, these results have implications to temporal networks. In this case it is
not that our knowledge is lacking about the network, but that the links change
as time progresses \cite{holme-temporal}. Social networks often display this sort of time dependence
\cite{social-dynamical} and many neural networks also change through time
\cite{neural-neworks-dynamics2, neural-networks-dynamics}. For these networks
caution must be taken before applying methods of extracting skeletons or
backbones since their changing states will lead to different results.

\section*{Acknowledgments}
We thank the Volkswagen Foundation for support.

\singlespacing
\bibliographystyle{unsrt}
\bibliography{paper}

\end{document}